\documentclass[a4paper,11pt]{article}
\usepackage{jcappub}
\usepackage{graphics,graphicx, color}
\usepackage{dcolumn, amsmath,amssymb}

\def\beq{\begin{equation}}
\def\eeq{\end{equation}}
\def\bey{\begin{eqnarray}}
\def\eey{\end{eqnarray}}

\title{Identifying Ultrahigh-Energy Cosmic-Ray Accelerators with Future Ultrahigh-Energy Neutrino Detectors}

\author[a,b]{Ke Fang}
\author[c,d]{Kumiko Kotera}
\author[a,b]{M. Coleman Miller}
\author[e,f]{Kohta Murase}
\author[e]{Foteini Oikonomou}
\affiliation[a]{Department of Astronomy, University of Maryland, College Park, MD, 20742-2421, USA}
\affiliation[b]{Joint Space-Science Institute, College Park, MD, 20742-2421}
\affiliation[c]{Sorbonne Universit\'es, UPMC Univ. Paris 6 et CNRS, UMR 7095,  Institut d'Astrophysique de Paris, 98 bis bd Arago, 75014 Paris, France}
\affiliation[d]{Laboratoire AIM-Paris-Saclay, CEA/DSM/IRFU, CNRS, Universite Paris Diderot,  F-91191 Gif-sur-Yvette, France}
\affiliation[e]{Department of Physics; Department of Astronomy \& Astrophysics; Center for Particle and Gravitational Astrophysics, The Pennsylvania State University, PA 16802, USA}
\affiliation[f]{Yukawa Institute for Theoretical Physics, Kyoto University, Kyoto 606-8502, Japan}
 
\abstract{
The detection of ultrahigh-energy (UHE) neutrino sources would contribute significantly to solving the decades-old mystery of the origin of the highest-energy cosmic rays. We investigate the ability of a future UHE neutrino detector to identify the brightest neutrino point sources, by exploring the parameter space of the total number of observed events and the angular resolution of the detector. 
The favored parameter region can be translated to requirements for the effective area, sky coverage and angular resolution of future detectors, for a given source number density and evolution history.  Moreover, by studying the typical distance to sources that are expected to emit more than one event for a given diffuse neutrino flux, we find that a significant fraction of the identifiable UHE neutrino sources may be located in the nearby Universe if the source number density is above $\sim10^{-6}\,\rm Mpc^{-3}$. If sources are powerful and rare enough, as predicted in blazar scenarios, they can first be detected at distant locations.
Our result also suggests that if UHE cosmic-ray accelerators are neither beamed nor transients, it will be possible to associate the detected UHE neutrino sources with nearby UHE cosmic-ray and gamma-ray sources, and that they may also be observed using other messengers, including ones with limited horizons such as TeV gamma rays, UHE gamma rays and cosmic rays. 
We find that for a $\gtrsim5\sigma$ detection of UHE neutrino sources with a uniform density, $n_s\sim{10}^{-7}-{10}^{-5}~{\rm Mpc}^{-3}$, at least $\sim100-1000$ events and sub-degree angular resolution are needed, and the results depend on the source evolution model.
}

\begin{document}

\maketitle

\section{Introduction}

The first detections of TeV - PeV ($10^{12}-10^{15}\,$eV) neutrinos by the IceCube collaboration \citep{Aartsen:2013bka, Aartsen:2013jdh, Aartsen:2014gkd, 2015PhRvD..91b2001A,Aartsen:2015ita,Aartsen:2015rwa} opened up the era of high-energy neutrino astronomy (see \cite{Waxman:2013zda,Meszaros:2014tta,Murase:2014tsa,Ahlers15} for reviews).  The energies above a few PeV are still uncharted territory, but the existence of EeV ($10^{18}\,$eV) neutrinos is guaranteed because they will be produced by the interaction of ultrahigh-energy cosmic rays (UHECR, charged particles with energies greater than $10^{18.5}$ eV) directly in the source environment, or during their propagation in the intergalactic medium. 
While the flux level of neutrinos produced at the source depends on the data and modeling, for the neutrinos generated during the propagation of UHECRs in the intergalactic medium (the so-called {\it cosmogenic} neutrinos) predictions are less uncertain although the chemical composition has a significant effect on the expected neutrino flux \cite{KAO10}. 

UHECRs were first detected decades ago, but their sources remain unknown. One reason is that the trajectories of these charged particles are bent by extragalactic and Galactic magnetic fields, and are thus difficult to trace back. Neutrinos, in contrast, propagate over cosmological distances under the influence of only gravity. Neutrinos produced by interactions of ions with matter or radiation are expected to have energies $\sim 3-5$\% of the original hadron energy \citep{2009herb.book.....D}, which means that EeV neutrinos will be unambiguous probes of the accelerators of UHECRs \footnote{Larger statistics at TeV-PeV energies could lead to a sooner discovery of point sources in this energy range, which however, are not guaranteed to be UHECR sources.}.  

Nonetheless, detecting an ultrahigh-energy (UHE) neutrino does not guarantee the identification of an UHECR ``accelerator''. Since there is no horizon for high-energy neutrino propagation (unlike for gamma-rays or cosmic-rays), it is likely that a range of astrophysical objects will lie  within the solid angle associated with a detected neutrino, making it difficult to determine the actual source.  Source catalogs obtained by electromagnetic observations could be helpful, but high-energy catalogs are usually incomplete due to flux limitations. Besides, a source association may be difficult if  sources of high-energy neutrinos are different from that observed in other wavebands, or if the events come from multiple types of sources.  A more robust way of finding the UHE sources would be to identify the bright sources on top of a diffuse background. 
As in traditional astronomy, implications of a point-source search have been studied in the literature of neutrino astronomy \citep{Lipari:2008zf,Silvestri:2009xb,Murase:2012df,2014PhRvD..90d3005A, 2016arXiv160701601M}.  
Importantly, in the UHE range, the atmospheric neutrino background can be safely neglected.  Thus, the success of such a search depends on the angular resolution and sensitivity of the detectors, but also on the source population  luminosity density (or energy budget) and emissivity history. 

Many existing and projected experiments have been proposed to detect EeV neutrinos, including the Antarctic Impulsive Transient Antenna (ANITA\citep{2010arXiv1003.2961T}) , the ANTARES telescope \citep{2011NIMPA.656...11A}, the Askaryan Radio Array (ARA \citep{2012APh....35..457A}), the Antarctic Ross Ice-Shelf ANtenna Neutrino Array (ARIANNA \cite{Barwick:2006tg}),  the Cubic Kilometre Neutrino Telescope (KM3NeT \citep{2016JPhG...43h4001A}),   the ExaVolt Antenna (EVA \cite{2011APh....35..242G}), the Giant Radio Array for Neutrino Detection (GRAND \cite{Martineau-Huynh:2015hae}), IceCube \cite{IceCube10}, IceCube-Gen2 \citep{2014arXiv1412.5106I}, the JEM-EUSO Mission \citep{2013arXiv1307.7071T}, Cherenkov Telescope Array (CTA \cite{Gora:2016mmy}), CHerenkov from Astrophysical Neutrinos Telescope (CHANT \cite{Neronov:2016zou}), and Neutrino Telescope Array (NTA \cite{2014arXiv1408.6244S}). These experiments employ very different techniques, and have a wide range of effective areas, angular resolutions, and detection efficiencies \citep{2016arXiv160708232C, 2016arXiv160708781S}.

In this work, we investigate the requirements for a future EeV neutrino detector to identify an UHE neutrino source. In Section~\ref{sec:sigma} we explore the prospects for pinpointing sources using detectors with a range of angular resolutions and effective areas. In Section~\ref{sec:UHECR} we examine the possibility of association between UHE neutrinos and UHECR sources.  We conclude in Section~\ref{sec:discussion} with a discussion of the impact of our results in light of existing experiments.  Finally, we point out that our work can also apply to neutrinos in the energy range of 100 TeV - EeV where the effect of atmospheric neutrinos is negligible.

\section{Capability of Point-source Detection}\label{sec:sigma}

The non-detection of UHE neutrinos using the IceCube Observatory and the Auger Observatory \citep{Aab:2015kma,Aartsen:2016ngq} sets an upper limit on the all-flavor flux: $E_\nu^2\Phi_\nu \lesssim 3\times10^{-8}\,\rm GeV\,cm^{-2}\,s^{-1}\,sr^{-1}$ at 1 EeV. 
Let us assume that a neutrino detector is sensitive to one flavor (e.g., $\nu_\tau$)~\footnote{Note that in practice a detector may be sensitive to more than one flavor.}. Then, for an operation time $T_{\rm obs}$, a detector with a ``neutrino'' effective area $A_{\rm eff}$ and an instantaneous fractional sky coverage  $f_{\rm cov}$ is expected to observe a total number of neutrinos (``events") 
\bey
\label{eq:Nev}
N_{\rm tot}^{\rm ev}(E) &\equiv&\int_E dE_\nu \, \Phi_{\nu_i}(E_\nu) \,A_{\rm eff}(E_\nu)\,T_{\rm obs}\,4\pi\,f_{\rm cov}.
\eey
Note that we have used the neutrino effective area, and a fraction of the neutrino energy is usually measured in detectors. Assuming the effective area does not strongly increase with energy in this energy range, $E\sim\Delta E\,\sim1$~EeV leads to  
\bey
\label{eq:Nev}
N_{\rm tot}^{\rm ev}(E) &\sim&\Phi_{\nu_i}(E) \,A_{\rm eff}(E)\,T_{\rm obs}\,4\pi\,f_{\rm cov}\, \Delta E.
\eey
Then, the IceCube and Auger limits imply
\bey
N_{\rm tot}^{\rm ev} (1\,{\rm EeV})&\lesssim& 990\,\left(\frac{A_{\rm eff}(1~{\rm EeV})}{10^{11}\,\rm cm^2}\right)\left(\frac{T_{\rm obs}}{5\,\rm yr}\right)\left(\frac{f_{\rm cov}}{0.5}\right)\ .
\eey

The expected number of events $N_{\rm tot}^{\rm ev}$ scales with the effective area of the detector, and could span a large range due to uncertainty about the sources of UHECRs. For reference, at 1 EeV, the effective area of some projected and future EeV neutrino experiments  ranges from $A_{\rm eff}\sim 10^9\,\rm cm^2$ (e.g., ARA-37 \cite{2015arXiv150708991A}) to $\sim 10^{11}\,\rm cm^2$ (e.g., GRAND \cite{Martineau-Huynh:2015hae}).

An EeV neutrino flux of $\sim 10^{-8}\,{\rm GeV\,cm^{-2}\,s^{-1}\,sr^{-1}}$ is comparable to the Waxman-Bahcall bound flux~\cite{Waxman:1998yy} as well as IceCube's diffuse neutrino flux, so it can be seen as a benchmark flux level that one could expect for a reasonable UHECR source scenario. It corresponds to $\sim 1000$ events for the ambitious detector parameters given above. 

Another relevant detector property is the angular resolution $\triangle \theta$, which plays a crucial role in pinpointing the sources. Below we leave $N_{\rm tot}^{\rm ev}$ and $\triangle \theta$ as free parameters, and investigate the source-search capability of a detector in the parameter space of $N_{\rm tot}^{\rm ev}$ and $\triangle \theta$. \\

\subsection{Theoretical Perspectives}
The most secure source of astrophysical EeV neutrinos is believed to be cosmogenic neutrino production by UHECRs interacting with the cosmic microwave background \citep{1987pcrp.book...99B, 1993PThPh..89..833Y, 2001PhRvD..64i3010E, 2009APh....31..201T}.
For a proton-dominated composition, including Galactic mixed composition, and a source emissivity evolution similar to the star-formation rate (SFR), the predicted fluxes of cosmogenic neutrinos lie within the narrow range of $E^2\Phi_{\nu}\sim 0.75-1.5\times 10^{-8}\,{\rm GeV\,cm^{-2}\,s^{-1}\,sr^{-1}}$ (that is comparable to the Waxman-Bahcall bound for a flat energy spectrum) around 1~EeV for 3 neutrino flavors, for a broad set of standard astrophysical parameters \cite{KAO10}. Indeed, the level of neutrino flux at these energies is mostly governed by the well-measured UHECR flux and by the source emissivity evolution up to redshift $\sim 2$, which is likely to follow roughly the history of star formation.

On the other hand, for a composition dominated by heavy nuclei such as iron, the predicted fluxes of cosmogenic neutrinos are significantly lower \citep{2008APh....29....1A, 2012PhRvD..86h3010A}, around $E^2\Phi_{\nu}\sim 10^{-9}\,{\rm GeV\,cm^{-2}\,s^{-1}\,sr^{-1}}$, following the nucleus-survival bound for a flat energy spectrum \cite{Murase:2010gj}. Such pessimistic cases are more difficult to test, and ultimately large neutrino detectors are required. 

UHE neutrinos are also expected to be produced at the source when UHECRs interact with ambient radiation and/or matter. Among steady UHECR sources, the most popular candidate sources have been active galactic nuclei. In particular, one of the most promising acceleration sites is the inner jet region of blazars, and blazars have also been considered as neutrino sources \cite{Mannheim:1995mm,Atoyan:2001ey,Atoyan:2002gu} (see also a review \cite{Murase:2015ndr} and references therein).  
Powerful blazars including quasar-hosted blazars (QHBs) and low-frequency peaked BL Lac objects may be UHECR accelerators, so that EeV neutrino production has also been expected both in the leptonic model \cite{Murase:2014foa} and the lepto-hadronic model \cite{Petropoulou:2016ujj,Padovani:2015mba}. The expected fluxes are $E^2\Phi_{\nu}\sim 10^{-8}-10^{-7}\,{\rm GeV\,cm^{-2}\,s^{-1}\,sr^{-1}}$, and optimistic models have already been ruled out by observations with e.g., IceCube and the Pierre Auger Observatory \citep{Aab:2015kma,Aartsen:2016ngq}. 
Other promising steady UHE neutrino sources include galaxy clusters and groups, which have been predicted to be PeV neutrino sources \cite{Murase:2008yt,Kotera:2009ms} and have also been suggested as the origin of IceCube's neutrinos \cite{Murase:2013rfa,Fang:2016amf}. The model predicts that low-energy neutrinos have a hard spectrum due to CR confinement whereas high-energy neutrinos have a steep spectrum due to CR escape. 
As shown in Refs.~\cite{DeMarco:2005va,Kotera:2009ms}, EeV neutrinos are mainly produced by photomeson production interactions with the cosmic infrared background in clusters, and the predicted flux is $E^2\Phi_{\nu}\sim 10^{-9}-10^{-8}\,{\rm GeV\,cm^{-2}\,s^{-1}\,sr^{-1}}$. 

A successful detection of UHE neutrinos would enable us to identify UHECR accelerators, and thus multiplet searches have been performed in the context of UHECR astronomy. The number of event clusters from UHECR accelerators, with arrival directions separated by less than a few degrees, that may be associated with a single source, constrains the apparent density of sources of UHECRs \citep{2009APh....30..306T, 2006JCAP...01..009C}.
If there is not an excess of UHECR multiplets beyond what is expected for an isotropic distribution of neutrinos, then a lower limit on the apparent number density of UHECR sources can be derived \cite{Takami:2014zva, 2009APh....30..306T, 2008JCAP...05..006K}. In Ref.~\cite{AugerBound} the apparent, local UHECR number density was shown to be consistent with $n_{s}^{\rm cr} \geq (0.06-5)\times10^{-4}\,\rm Mpc^{-3}$. The corresponding average UHECR luminosity is $EL_E^{\rm CR}\lesssim{10}^{40}-{10}^{41.5}~{\rm erg}~{\rm s}^{-1}$. 
Thus, if UHECRs are steady and isotropically emitted from their sources, the number density and luminosity  of the associated UHE neutrinos should follow the same constraints.

However, the above scenario is not necessarily true. In general, a population that dominates the observed UHE neutrino sky need not also dominate the observed UHECR sky. This is because neutrinos mainly come from distant sources whereas UHECRs mainly come from local sources. As a specific example, for beamed sources such as blazars, the apparent number density of UHECR accelerators (measured by neutrinos), $n_s$, can naturally be much smaller than the apparent number density of UHECR sources (measured by UHECRs), $n_s^{\rm cr}$ (see discussion in Ref.~\cite{Murase:2011cy}).  
If acceleration regions (e.g., inner jets) are relativistically boosted, relativistic beaming causes particles to be emitted within a narrow cone.  However, charged particles should be significantly isotropised by intervening magnetic fields, except for rare sources residing in cosmic voids with weak magnetic fields \cite{Murase:2011cy}.  As a result, the apparent source number density of UHECR sources can be significantly larger than $n_{s}$, and may even be comparable to the ``true'' source density $n_s^{\rm true}$. For beaming sources, we must consider lower source number densities for UHE neutrinos compared to ones for UHECR sources if the parent UHECRs are significantly isotropised. To represent this situation we therefore consider $n_s=10^{-7}\,\rm Mpc^{-3}$ in the Case~II and Case~III in Sec~\ref{sec:sigma}. This number density is comparable to the {\it total} number density of {\it Fermi} blazars, which have typical jet opening angles of a few degrees \citep{2012ApJ...751..108A, Ajello:2013lka}. 
Note that our choice is quite conservative in the context of  the number of available  sources in the sky.  In realistic models, the effective source number density, which is calculated based on the luminosity function, is significantly lower than the total number density \cite{2016arXiv160701601M}. For example, the effective source number density of BL Lac objects is $n_s\sim{10}^{-9}-{10}^{-8}~{\rm Mpc}^{-3}$.  Also, QHBs are rarer but more powerful, so that they are more efficient and powerful neutrino sources \cite{Murase:2014foa,Dermer:2014vaa}.  The total number density of QHBs is $n_s\sim{10}^{-9}~{\rm Mpc}^{-3}$ at $z=0$ but the effective source number density is as small as $n_s\sim{10}^{-12}-{10}^{-11}~{\rm Mpc}^{-3}$ (with a redshift evolution stronger than the SFR) although QHBs show a strong redshift evolution and high-redshift contributions are more important than usual.  
Remarkably, Ref.~\cite{Murase:2014foa} predicted that cross-correlation signals with {\it Fermi} blazars can be detected because most of the diffuse neutrino flux is dominated by luminous blazars.

\subsection{The Calculation Method and Results}\label{sec:sigma}
 To assess the capability of detectors to find point sources against a diffuse background, we use the statistical tool described in Ref.~\cite{Fang:2016hyv}. A data set of $N$ detected neutrinos contains $N (N - 1)/2$ unique pairs. Using the angular separation ${\alpha}_{ij}$ between each pair of events $i$ and $j$, we construct an unbinned likelihood 
\beq
\ln{\cal L}(f )=\sum_{i<j} \ln\left[f \,{\mathcal A}_{\rm point}({\alpha}_{ij})+(1-f )\,{\mathcal A}_{\rm diff}({\alpha}_{ij})
\right]\ .
\eeq 
Here  ${\mathcal A}_{\rm point}$ and ${\mathcal A}_{\rm diff}$ correspond to the  probabilities of having an angular separation $\alpha_{ij}$\footnote{Because we consider a uniform angular resolution in this work, $\alpha_{\ij}$ is equivalent to  the $\bar{\alpha}_{ij}$ in Ref.~\cite{Fang:2016hyv}.} for, respectively, an individual point source and for isotropic diffuse sources. We maximize the likelihood over $f$, which is the fraction of the pairs that share the same direction (same-source pairs).  To evaluate the significance of the signal, we introduce a test statistic (TS) defined as
\beq
{\rm TS}  = 2 \ln \left[ \frac{{\cal L} (\hat{f})}{{\cal L}(f=0)} \right]\,
\eeq
where ${\hat f}$ is $f$ that maximizes the likelihood function.

For a given dataset of size $N^{\rm ev}_{\rm tot}$ and detector angular resolution $\triangle \theta$, we generate a large number of synthetic reference datasets from an isotropic background. The percentile of the TS of the data out of the TS of the references determines the confidence level at which we reject the null hypothesis of no individual point sources (that is, the $p$-value). We can, equivalently, quote the corresponding number of standard deviations for this confidence level for a Gaussian distribution.

For simplicity, we  assume that the detector in consideration has a uniform sensitivity and a uniform angular resolution  over the entire sky ($f_{\rm cov}=1$). This setup can be easily adapted to more realistic sensitivity and angular resolution maps. We consider 8 different values of $\triangle\theta$ ranging from $0.05^\circ$ to $3^\circ$, and 10 different values of  $N^{\rm ev}_{\rm tot}$ with ranges  depending on the source number density  (from 50 to 3000 Mpc for $n_s=10^{-5}\,\rm Mpc^{-3}$ (uniform) and $10^{-7}\,\rm Mpc^{-3}$ (SFR), and from 10 to 500 Mpc for $n_s=10^{-7}\,\rm Mpc^{-3}$ (uniform) and $10^{-9}\,\rm Mpc^{-3}$ (SFR)). For each set of $\left(\Delta\theta, N_{\rm tot}^{\rm ev}\right)$, we generate $10^5$ synthetic reference datasets from an isotropic background. We also perform $10^3$ tests using data generated with point sources. Finally, we use the average $p$-value of all $10^3$ tests to determine the expected significance of detection. Note that real data has statistical fluctuations and thus does not necessarily result in the mean value predicted here.   

We generate the mock data by drawing events randomly from the background or the sources, and then smoothing by the point-spread function  (PSF) at the injection direction. We draw the total number of sources from a Poisson distribution with a mean determined by the source distribution and  the source volume.  We assume that all the point sources have the same luminosity, and consider four scenarios of source distributions: 
\begin{enumerate}
\item I)  a uniform number density $n_s = 10^{-5}\,\rm Mpc^{-3}$ up to a sharp edge at 2~Gpc; 
\item II) a uniform  number density $n_s = 10^{-7}\,\rm Mpc^{-3}$  up to a sharp edge at 2~Gpc; 
\item III) a number density that is $n_s = 10^{-7}\,\rm Mpc^{-3}$ locally but that is proportional to the SFR up to redshift $z_{\rm max}=6$;
\item  IV) a number density that is $n_s = 10^{-9}\,\rm Mpc^{-3}$ locally but that is proportional to the SFR up to redshift $z_{\rm max}=6$. 
\end{enumerate}

Our pick of a uniform distribution with a cutoff at 2 Gpc approximates source distributions in the relatively nearby universe, and is computationally efficient for the large number of  realizations required in this work.  It is significantly more favorable to source detection than is the scenario in which sources follow the SFR, as is seen by a comparison of the cases II and III in Figure~\ref{fig:sigma7}. 

\begin{figure}[!htb]
\centering
\epsfig{file=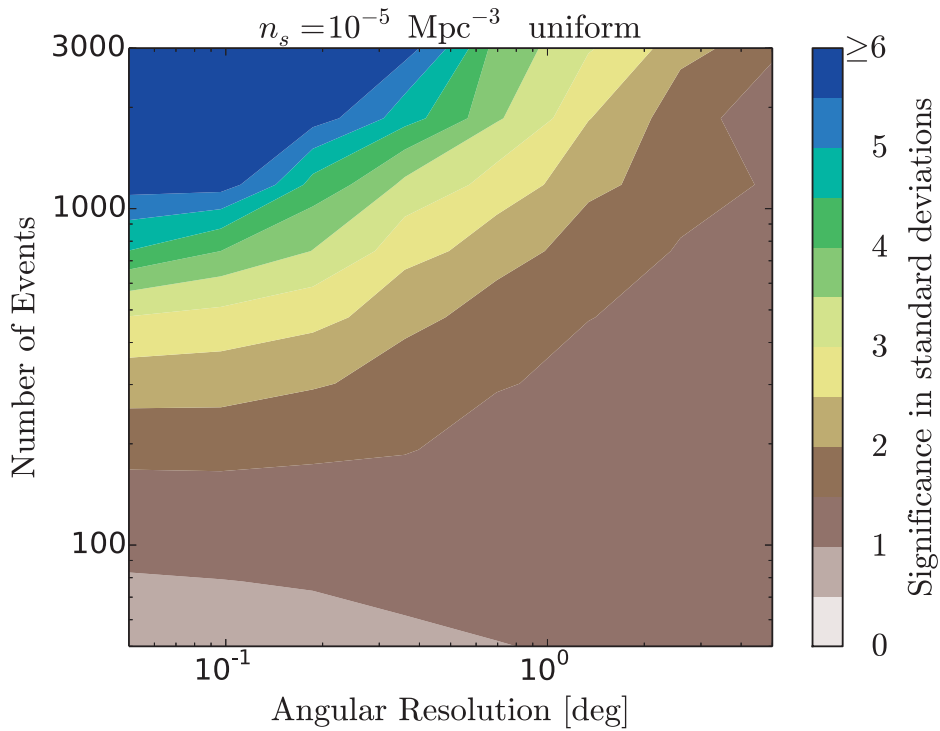,width=0.9\textwidth}
\caption{Significance of detection of point sources of UHE neutrinos by experiments with various angular resolutions and numbers of detected events. The color coding corresponds to the confidence level to reject an isotropic background using the statistical method from Ref. \cite{Fang:2016hyv}. We assume that all of the sources have the same luminosity, and that the sources follow a uniform distribution with a number density $10^{-5}\,\rm Mpc^{-3}$ up to 2 Gpc (case I). With this source number density, $\sim1000$ events and $\sim 0.1^\circ$ angular resolution are needed to reach a 5$\sigma$ detection of point sources. In the above calculation, $f_{\rm cov}=1$ is used; fewer events are required {\it in the field of view} if $f_{\rm cov}$ is smaller.}
\label{fig:sigma}
\end{figure}

\begin{figure}[!htb]
\centering
\epsfig{file=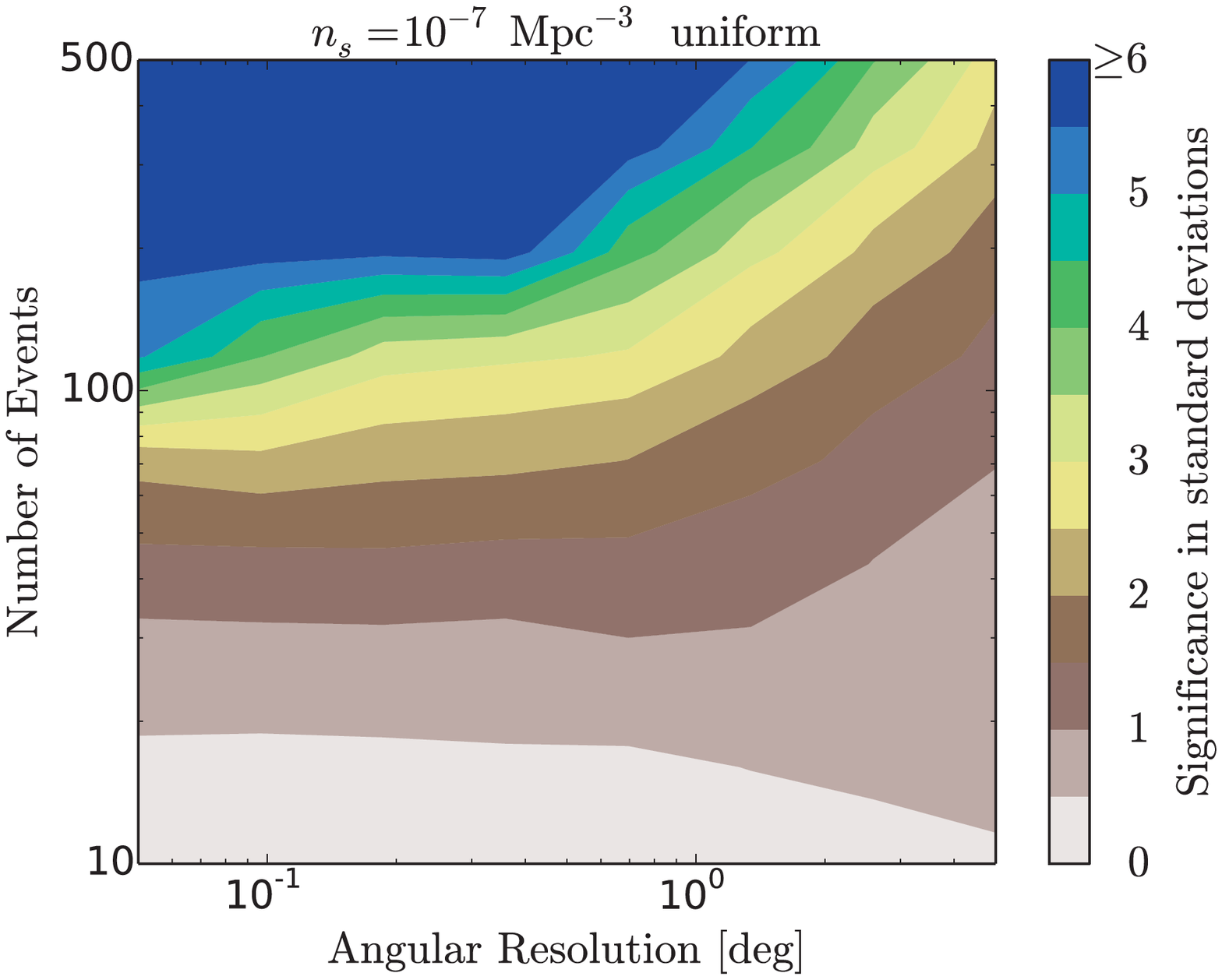,width=0.54\textwidth}
\epsfig{file=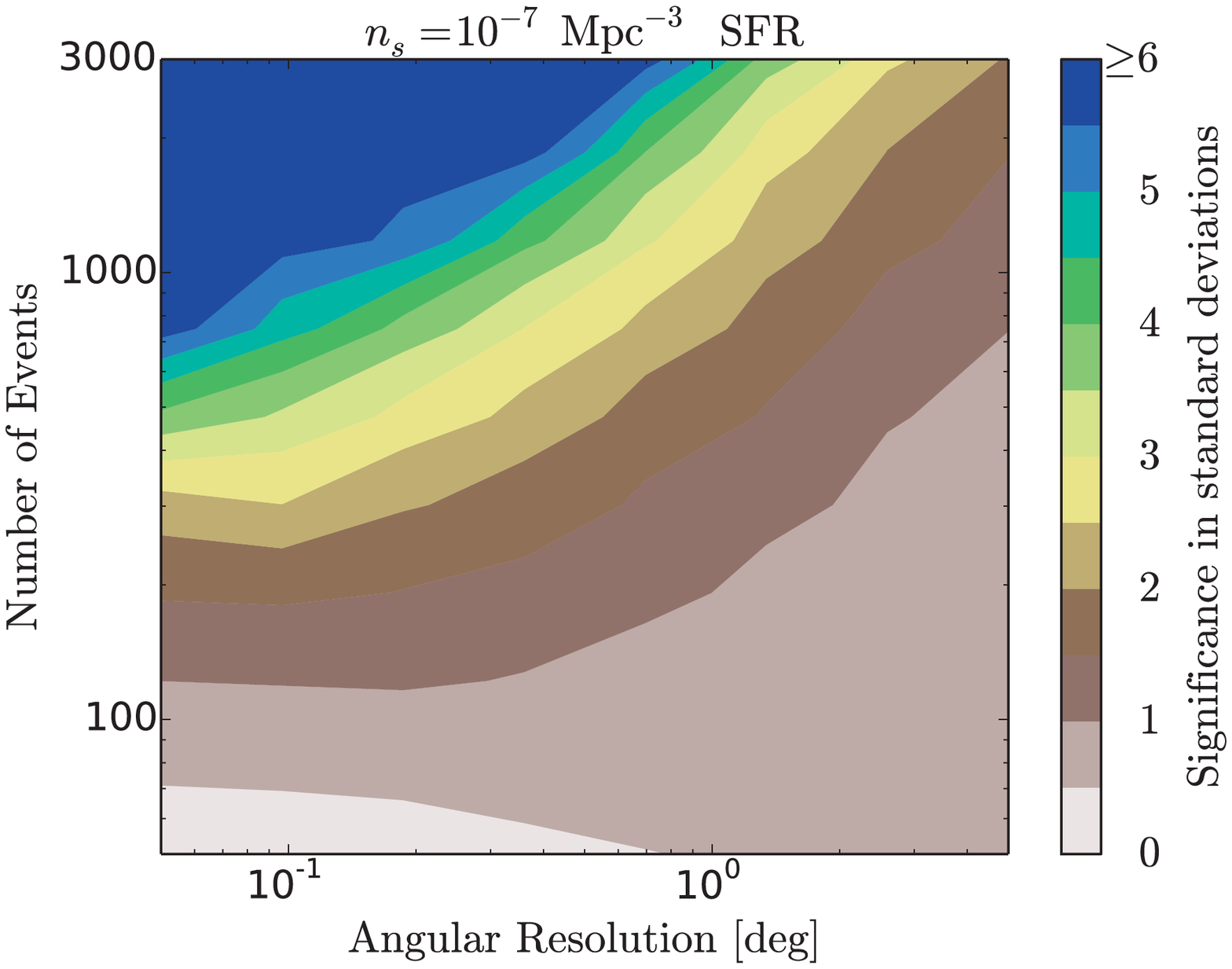,width=0.54\textwidth}
\epsfig{file=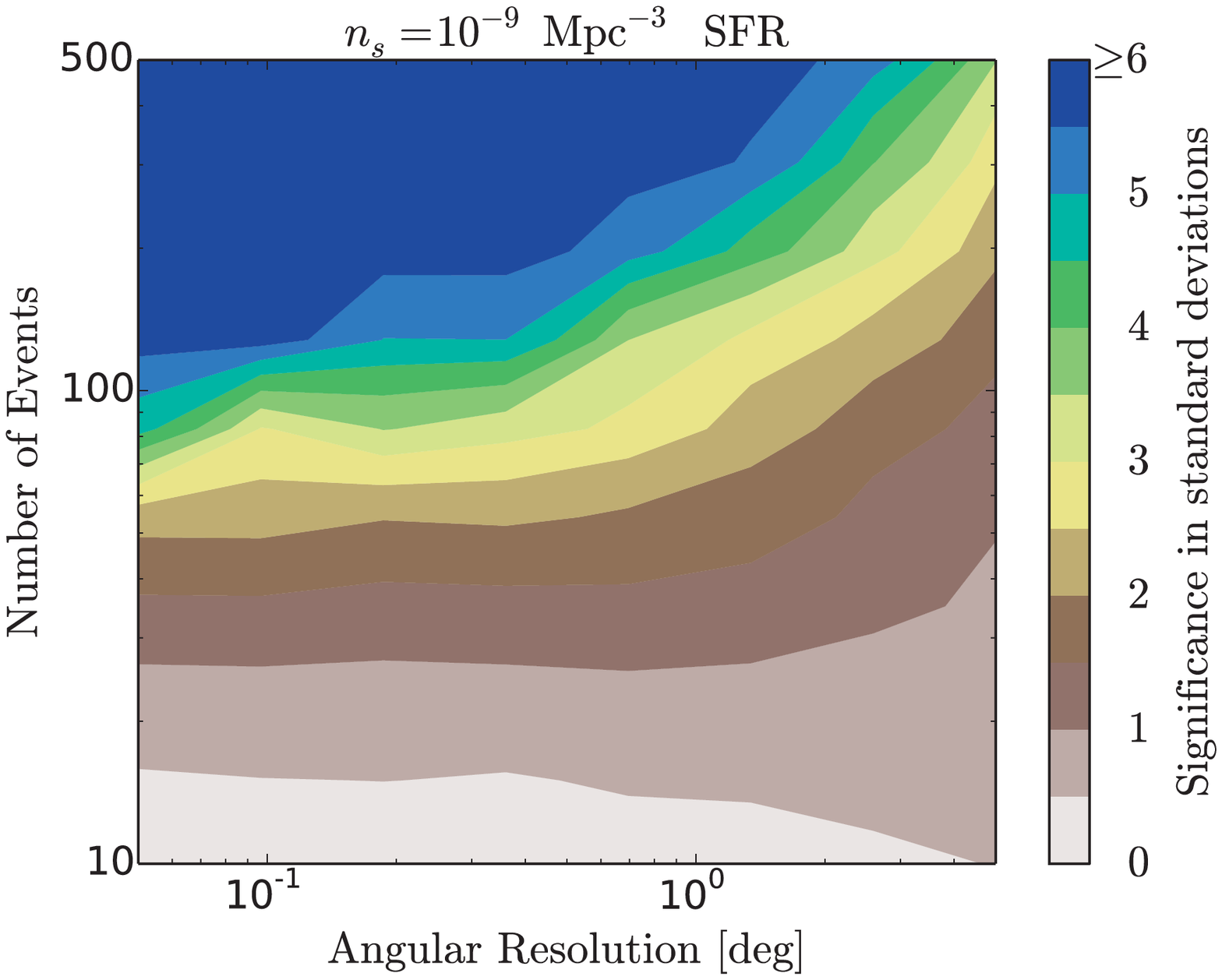,width=0.54\textwidth}
\caption{The same as Figure~\ref{fig:sigma}, but assuming that sources  follow a uniform distribution with a number density $10^{-7}\,\rm Mpc^{-3}$ up to 2 Gpc (top, case II), or a number density that is $10^{-7}\,\rm Mpc^{-3}$  (middle, case III) or $10^{-9}\,\rm Mpc^{-3}$  (bottom, case IV) locally but that is proportional to the SFR up to redshift $z_{\rm max} =6$. 
In general to reach the same significance level of detection, more events will be needed if sources have a larger source number density, or if more sources are distributed at large distances. 
} 
\label{fig:sigma7}
\end{figure}

Figures~\ref{fig:sigma} and~\ref{fig:sigma7} show the significance of point-source detection by a detector in our parameter space of event numbers and angular resolution. For sources with $n_{\rm s}= 10^{-5}\,\rm Mpc^{-3}$ (case I), a 3$\sigma$ detection requires at least 500 events with $\Delta\theta\sim0.1^\circ$, roughly 1000 events with $\Delta\theta\sim 0.5^\circ$, and a greater number of events with detectors that have an angular resolution poorer than a degree. A 5$\sigma$ detection generally requires a few thousand events and an angular resolution better than $0.5^\circ$. At a fixed significance level, the required number of events is roughly independent of angular resolution for $\Delta\theta \lesssim 0.1^\circ$ at 5$\sigma$, but increases notably for angular resolutions worse than a few tenths of a degree. This change happens when the chance of getting background events from adjacent sources due to the poor PSF becomes considerable, that is, the number of false point sources in the background is not negligible (see Section IV of Ref.~\cite{2016arXiv160701601M} and considerations in Ref.~\cite{Murase_Takami09,Takami:2011nn} for constraints on UHECR sources).  
We confirmed that our results agree well with calculations based on multiplet analyses performed by Refs.~\cite{2014PhRvD..90d3005A, 2016arXiv160701601M}. A $1.6\sigma$ limit corresponds to $N_{\rm tot}^{\rm ev}\sim200$, which is consistent with the six-year lower limit on the number density $n_s\gtrsim{10}^{-5}~{\rm Mpc}^{-3}$ for no redshift evolution and $f_{\rm cov}=0.5$ \cite{2016arXiv160701601M}.  
Note that alternate point-source detection methods, such as standard autocorrelation methods or the method of Ref.~\cite{Braun}, would require more events and/or better angular resolution (see the discussion in Ref.~\cite{Fang:2016hyv}).

\begin{figure}[!htb]
\centering
\epsfig{file=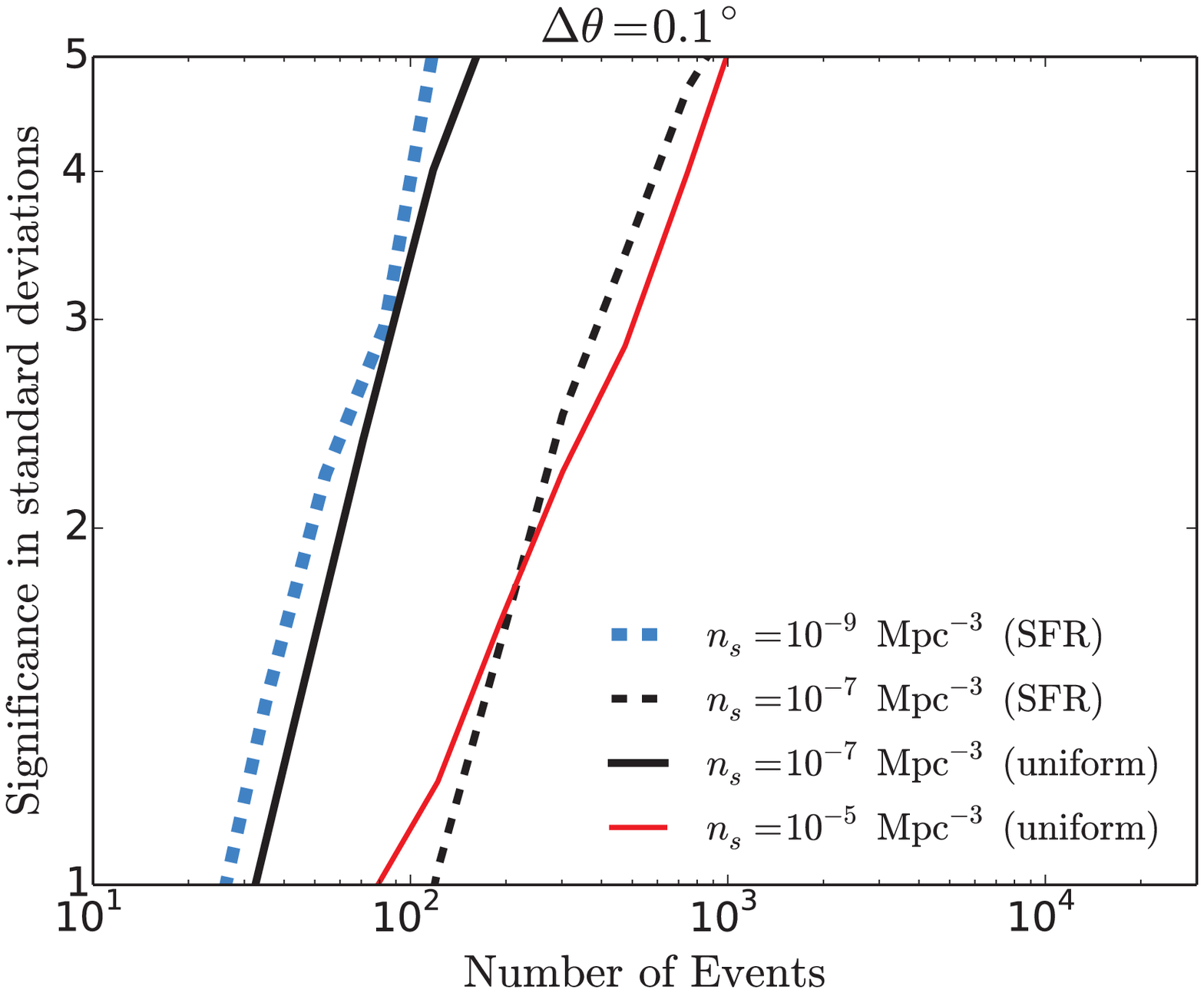,width=0.85\textwidth}
\caption{Significance of detection of point sources as a function of  numbers of detected events, for   the $n_s$ and redshift evolution models considered in Figs.~\ref{fig:sigma} and \ref{fig:sigma7}, taking a uniform angular resolution $\Delta\theta=0.1^\circ$ and assuming a uniform sensitivity over the entire sky.   This figure illustrates how $N^{\rm ev}_{\rm tot}$ varies as a function of the Gaussian significance $\sigma$, and that $\gtrsim100-1000$ events are needed for a significant detection in these cases.}
\label{fig:sigma_N}
\end{figure}

The significance of point source detection depends on the source number density as well as the source evolution model. In general, to reach a given confidence level, more events will be needed if the total number of sources is larger or if the sources lie at greater distances. For example, if sources follow a uniform distribution with $n_{\rm s}=10^{-4}\,\rm Mpc^{-3}$ up to a sharp edge at 2 Gpc, a 3$\sigma$ detection would require about 1700 events even with $0.1^\circ$ angular resolution. In contrast, the top panel of Figure~\ref{fig:sigma7} shows that with $n_s = 10^{-7}\,\rm Mpc^{-3}$ and a uniform distribution, the same level of detection can already be reached by 100 events with $1^\circ$ angular resolution.

To understand the impact from source evolution models, in the  middle and bottom panels of Figure~\ref{fig:sigma7} we show the significance of point-source detection assuming that sources follow the SFR (case III and IV). To model the star formation rate, we assume that the source number density scales with redshift $z$ as $(1+z)^{3.4}$ for $z<1$, $(1+z)^{-0.3}$ for $1<z<4$, and $(1+z)^{-3.5}$ for $z>4$ \citep{HB06}. With a significant source population distributed at large distances, an $5\sigma$ identification of  sources with a local density $n_s = 10^{-9}\,\rm Mpc^{-3}$ requires $\sim100$ events,  and that with $n_s  = 10^{-7}\,\rm Mpc^{-3}$ requires $\sim1000$.

Finally, we summarize the above four cases in Figure~\ref{fig:sigma_N}, by showing the significance of detection of point sources as a function of the total number of detected events at an angular resolution $\Delta\theta = 0.1^\circ$. Note that we have assumed $f_{\rm cov}=1$ in the above calculation. The required total number of events (in the field of view) would be less with a smaller value of $f_{\rm cov}$.  Also, even if $5\sigma$ detections require sufficient statistics, it is easier to find hints of UHE neutrino sources if the sources are rare. For example, for sources following a uniform distribution with $n_{\rm s}=10^{-7}\,\rm Mpc^{-3}$ and the SFR distribution with $n_{\rm s}=10^{-9}\,\rm Mpc^{-3}$, one can place $2\sigma$ limits only with dozens of neutrino events.

\section{Cumulative Contributions of the Brightest Sources}\label{sec:UHECR}
In principle, the point-source detection may be expected for powerful sources rather than nearby dim sources. For standard-candle sources with a given luminosity, one expects that the nearest source will be the easiest to detect as a single source. However, in the case of neutrinos, each of observed multiplets may consist of contributions from many neutrino sources located along the line of sight \cite{2016arXiv160701601M}. Although they are useful to place limits on the source number density \cite{2016arXiv160701601M}, for the purpose of detecting a single source, it is important to figure out cumulative contributions of the nearest sources, which we address in this section.  

EeV neutrinos would be direct probes of UHECR sources distributed up to cosmological distances. 
Bright sources such as blazars have been detected up to a few gigaparsecs. However, for dimmer sources, source catalogs may be incomplete beyond a few hundred megaparsecs because the catalogs are flux-limited and/or the field of view of detectors is small. At energies above $\sim 100$\,TeV, gamma-rays have a horizon of a few tens of Mpc. UHE photons above $\sim10^{19}$~eV have been proposed as a smoking gun of UHECR accelerators~\cite{Murase:2009ah}, and the effective energy-loss length is $\sim30-100$~Mpc since particle energy remains almost the same through cascades in the deep Klein-Nishina regime~\cite{Murase:2009ah,Murase:2011yw}.  
The observation of UHECRs are also limited by the so-called Greisen-Zatsepin-Kuzmin (GZK) horizon \cite{G66, ZK66}, due to interactions with the photons of the cosmic microwave background.  Although the majority of the EeV neutrino flux should come from distant sources around $z\sim1-2$ (see Eq.~\ref{eq:Nev} below), one expects that some bright neutrino sources lie in the nearby Universe, if neutrinos can be produced at the source.
We calculate in this section the typical distance to sources that can be identified using a UHE neutrino detector, in order to assess whether it is possible to associate such sources with nearby objects observed with other messengers.

The number of particles expected from one source with a neutrino emission rate per energy $dL_E/dE$ (where $L_E$ is the one-flavor neutrino luminosity per energy), located at distance $D$ is
\begin{equation}
N_{\rm 1s}^{\rm ev} (E) \sim \frac{1}{4\pi \, D^2}\frac{dL_E}{dE}\,A_{\rm eff}\,T_{\rm obs}\, \Delta E \ .
\end{equation}
For a neutrino source number density $n_{\rm s}$ at $z=0$, and assuming that all sources are identical (``standard candles''), the total expected number of detected events is
\bey
\label{eq:Nev}
N_{\rm tot}^{\rm ev} (E) &\sim& \xi_{z} n_{\rm s}  c\,t_{\rm H} \frac{d{L_E}}{dE} A_{\rm eff}\,T_{\rm obs}\,f_{\rm cov}  \Delta E \ ,
\eey
where $t_{\rm H}=\int_0^{z_{\rm max}}\,(dt/dz)\,dz$ is the Hubble time, $\xi_z=0.6$ if the source emissivity is independent of redshift (no evolution), and $\xi_z=2.5$ if the source emissivity follows the star formation rate, assuming that $\alpha\sim 2$ (see, e.g., \cite{Waxman:1998yy}).  
We can thus write
\begin{eqnarray}
{N_{\rm 1s}^{\rm ev}} &\sim&  \frac{{N_{\rm tot}^{\rm ev}}}{4\pi\,D^2n_{\rm s}c\,t_{\rm H}\xi_z f_{\rm cov}}\nonumber \\
&\simeq&1.6\, \left(\frac{N_{\rm tot}^{\rm ev}}{10^3}\right)  \left(\frac{D}{50\,{\rm Mpc}}\right)^{-2} \left(\frac{n_{\rm s}}{10^{-5}\,{\rm Mpc}^{-3}}\right)^{-1} \left(\frac{f_{\rm cov}}{0.9}\right)^{-1}  \left(\frac{\xi_z}{0.5}\right)^{-1}\ .
\end{eqnarray}
In this expression, all the single source parameters and the detector characteristics are encapsulated in the total number of events $N_{\rm tot}^{\rm ev}$. 

For a given neutrino luminosity, a single source is detected if $N_{\rm 1s}^{\rm ev}\gtrsim{\rm a~few}$ (implying the detection of multiplets due to a single source). By setting $N_{\rm 1s}^{\rm ev}=1$, one obtains the critical distance \cite{2016arXiv160701601M} \footnote{It essentially corresponds to the critical sample-variance distance defined by equating the flux of the brightest neutrino source (at which the number of sources becomes unity) to the point-source sensitivity \cite{Murase:2012df}. In the background free case, this critical distance $D_{\rm lim}$ is given by setting $N_{\rm 1s}^{\rm ev}=2.4$ for a 90\% CL sensitivity.}
\begin{eqnarray}\label{Dcri}
D_{\rm cri} &\equiv& \left(\frac{N_{\rm tot}^{\rm ev}}{n_{\rm s}c\,t_{\rm H}\xi_z f_{\rm cov}\,4\pi}\right)^{1/2}\\
&\simeq&63\,{\rm Mpc}\, \left(\frac{N_{\rm tot}^{\rm ev}}{10^3}\right)^{1/2}  \left(\frac{n_{\rm s}}{10^{-5}\,{\rm Mpc}^{-3}}\right)^{-1/2} \left(\frac{f_{\rm cov}}{0.9}\right)^{-1/2}  \left(\frac{\xi_z}{0.5}\right)^{-1/2},\ \nonumber
\end{eqnarray}
above which there is no detectable point source for a given luminosity.  
The critical distance can be compatible with cosmological distances of a few Gpc. For $n_s\gtrsim{10}^{-6}~{\rm Mpc}^{-3}$, one can see from this estimate that the detection of real multiplets will point to sources that are only in the local Universe, and in particular within the GZK horizon (of order $200\,{\rm Mpc}$ above a cosmic-ray energy of $60\,{\rm EeV}$). If UHECR accelerators are {\it neither beamed nor transient}, it would thus be possible to cross-correlate the position of UHE neutrinos with cosmic rays. 

Note that ${N_{\rm 1s}^{\rm ev}}$ cannot exceed the total number of events. ${N_{\rm 1s}^{\rm ev}} = N_{\rm tot}^{\rm ev}$ leads to a  distance below which the expected event from a single source saturates:
\begin{equation}
D_{\rm sat} \simeq 2\,  {\rm Mpc} \left(\frac{n_{\rm s}}{10^{-5}\,{\rm Mpc}^{-3}}\right)^{-1/2} \left(\frac{f_{\rm cov}}{0.9}\right)^{-1/2}  \left(\frac{\xi_z}{0.5}\right)^{-1/2}\ .
\end{equation}

Given a mean of $N_{\rm 1s}^{\rm ev}$, the probability of producing $m$ events is  \cite{2016arXiv160701601M} 
\beq
P(m|N_{\rm 1s}^{\rm ev}) =\frac{{e^{-N_{1s}^{\rm ev}}\,\left(N_{1s}^{\rm ev}\right)^m}/{m!}}{\sum_{m=0}^{N_{\rm tot}^{\rm ev}} {e^{-N_{1s}^{\rm ev}}\,\left(N_{1s}^{\rm ev}\right)^m}/{m!} }\,,
\eeq
considering that a source can emit events with a number between 0 and $N_{\rm tot}^{\rm ev}$ following a Poisson distribution. One recovers the well-known result $P(m|N_{\rm 1s}^{\rm ev}) =e^{-N_{1s}^{\rm ev}}\,\left(N_{1s}^{\rm ev}\right)^m/{m!}$ when the total number of events is not too small. 

The probability to find a source producing multiplets with order $M$ or higher is 
\beq \label{eq:Nmp_1s}
P_{1s}^{\rm mp} = \sum_{m\geq M} \,P(m|N_{\rm 1s}^{\rm ev}).
\eeq
Then, the number of neutrino sources producing $M$ or higher multiplets becomes \cite{2016arXiv160701601M}
\beq
{\mathcal N}_{s} = f_{\rm cov} \int_{D_{\rm min}}^{D_{\rm max}}P_{\rm 1s}^{\rm mp}\,n_s\,4\pi D'^2 dD'.
\eeq
The cumulative fraction of neutrino sources located within distance $D$ is given by
\beq\label{eqn:f_mp}
f_{\rm mp} = \frac{f_{\rm cov}}{{\mathcal N}_{s}}\, \int_{D_{\rm min}}^{D}P_{\rm 1s}^{\rm mp}\,n_s\, 4\pi D'^2 dD'\,.
\eeq

Next, we consider the number of pairs (i.e., the multiplets themselves) rather than the number of neutrino sources.  The expected value of the number of same-source pairs from a single source is
\beq \label{eq:Npair_1s}
N_{1s}^{\rm pair} =\sum_{m\geq2}\frac{m(m-1)}{2}\,P(m|N_{\rm 1s}^{\rm ev}) 
\eeq
where $m(m-1)/2$ is the number of pairs if a source emits $m$ events. The total number of pairs from all neutrino sources in the sky can be calculated as 
\beq
N_{\rm tot}^{\rm pair} = f_{\rm cov}\int_{D_{\rm min}}^{D_{\rm max}}N_{\rm 1s}^{\rm pair}\,n_s\,4\pi D'^2 dD',
\eeq
where $D_{\rm min}$ is the distance of the closest source, which is introduced to demonstrate effects of the sample variance around $D_{\rm cri}$ 
for the ensemble-averaged brightest (nearest) source. The cumulative fraction of pairs by neutrino sources located within distance $D$ is 
\beq\label{eqn:f_ss}
f_{\rm pair} = \frac{f_{\rm cov}}{N_{\rm tot}^{\rm pair}}\, \int_{D_{\rm min}}^{D}N_{\rm 1s}^{\rm pair}\,n_s\,4\pi D'^2 dD'.
\eeq
This quantity allows us to evaluate the relative contribution of nearby sources in the total number of pairs.  However, one should note that for given data this does not address whether higher multiplets can be discriminated from many doublets. For example, let us assume that four events are found in a sky region within the angular resolution. They may consist of a quartet by a single source or two doublets by two sources with different distances.   

\begin{figure}[!htb]
\centering
\epsfig{file=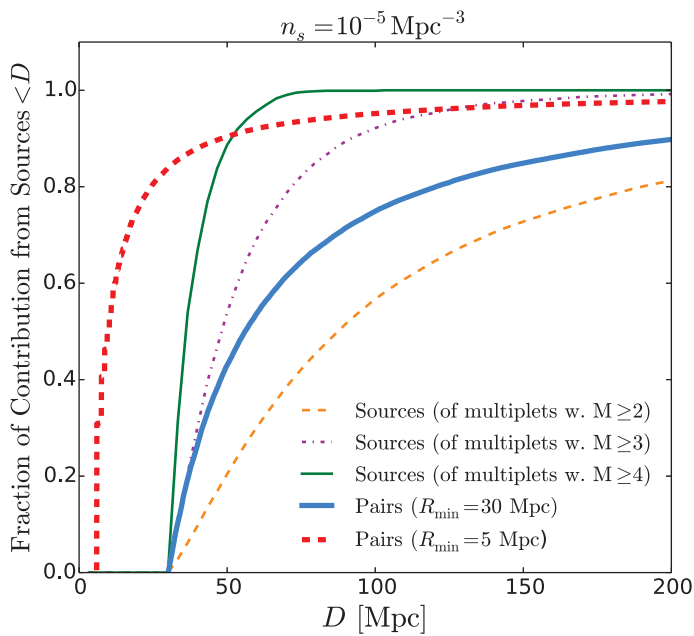,width=0.65\textwidth}
\caption{The chance, or the fraction of same-source pairs and multiplets with an order greater than 2, 3, and 4 being  produced by sources located within distance $D$. The source number density is fixed as $n_s=10^{-5}\,\rm Mpc^{-3}$. The minimal source distance is assumed to be $D_{\rm min}=5$ Mpc (dashed red line) and 30 Mpc (all the rest). The results were computed by numerical simulation, in which $\sim 10^4$ iterations (with $N_{\rm tot}^{\rm ev}=1000$) were performed and the locations of all occurrences of multiplets were recorded. The chance is determined by the fraction of the number of multiplets from sources within $D$ out of the total number of multiplets. The fractions can also be calculated using equation~\ref{eqn:f_ss} and ~\ref{eqn:f_mp}.
}
\label{fig:fracMultiplet}
\end{figure}

Figure~\ref{fig:fracMultiplet} presents the chance of contribution from sources within distance $D$ to multiplet events. Specifically, the chance is determined by the average fraction of the number of multiplets from sources within $D$ out of the total number of multiplets in many numerical realizations. 
We fix the local number density to be $n_s=10^{-5}\,\rm Mpc^{-3}$,  and set the minimum source distance $D_{\rm min}$ to either 5~Mpc or 30~Mpc  for demonstration purposes.
The minimum source distance sets the cutoff of the fraction $f_{\rm ss}$ and can influence strongly the value of $f_{\rm pair}$ within $D<D_{\rm cri}$.  As expected, $D_{\rm sat}$  and $D_{\rm cri}$ remain the principal parameters, which set the range of distances from which most real multiplets are expected. As $D_{\rm sat}$ does not depend on $N_{\rm tot}^{\rm ev}$, the fraction of the total contribution barely depends on the total number of events until distances around $D_{\rm cri}$. 
Importantly, the contribution of local neutrino sources is more prominent  for higher multiplets (compare cases with $M\geq2,3,4$). 

\begin{figure}[!htb]
\centering
\epsfig{file=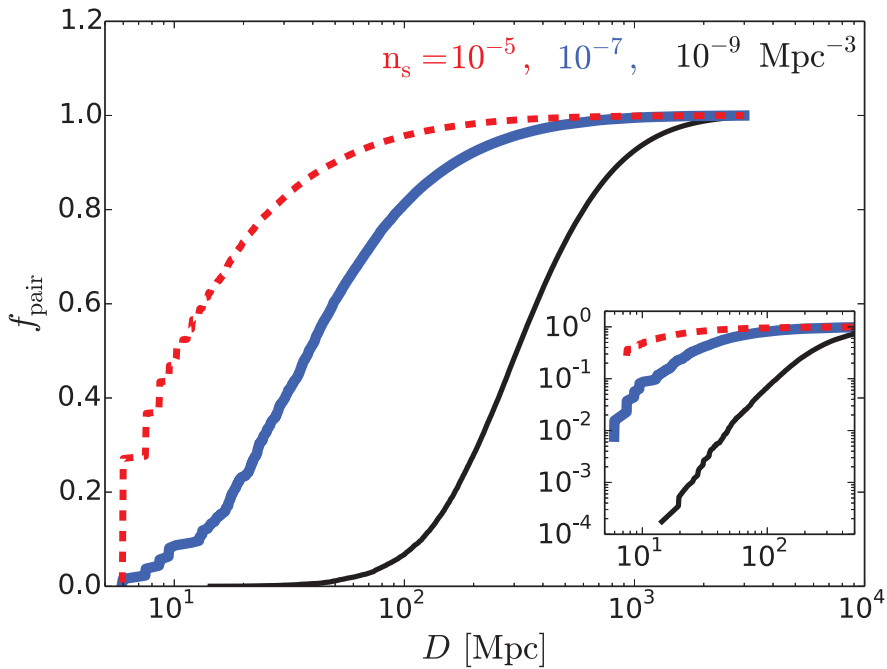,width=0.85\textwidth}
\caption{The chance of same-source pairs being produced by sources located within distance $D$, for a source number density $n_s = 10^{-5},\,10^{-7},\,10^{-9}\,\rm Mpc^{-3}$.  The minimal source distance is assumed to be $D_{\rm min}=5$ Mpc. Most pairs are expected to come from sources within $D_{\rm cri}$ (with $N_{\rm tot}^{\rm ev}=1000$). In addition, the subplot with a logarithmic y-axis shows that $f_{\rm pair} \propto D^3$ for $D\lesssim D_{\rm sat}$.
}
\label{fig:fracPairNs}
\end{figure}

Figure~\ref{fig:fracPairNs} presents the fraction of pairs from sources within $D$, for $n_s = 10^{-9},\, 10^{-7},$ and $10^{-5}\,\rm Mpc^{-3}$.
The minimum distance is set to be $R_{\rm min}=5\,\rm Mpc$ for all cases. The bulk of pairs are contributed from sources within $D_{\rm cri}$, which corresponds to an injection of $N_{\rm tot}^{\rm ev}=1000$.  In addition, the subplot in Figure~\ref{fig:fracPairNs} shows that $f_{\rm pair}\propto D^3$ for distances below $D_{\rm sat}$. This is because the expected value of the number of pairs from all potential sources within $D_{\rm sat}$ is limited to $N_{\rm tot}^{\rm ev}$, and the contribution simply scales to the volume of sources.  Also, effects of the cosmology as well as the redshift evolution become relevant at large distances although they are not shown in this figure~\footnote{Point-source limits are weaker for sufficiently rare sources because of effects of the cosmology~\cite{Murase:2012df,2016arXiv160701601M}.}.

The cumulative contribution to the number of sources or pairs is sensitive to the source number density. 
However, as inferred by Eq.~(\ref{Dcri}), the above figures demonstrate that, as long as the total number of events $N_{\rm tot}^{\rm ev}$ is small enough and/or the source number density $n_s$ is large enough, multiplets may originate mostly from nearby distances. For example, for $N_{\rm tot}^{\rm ev}\sim100$, most of the contributions come from sources at $\lesssim 200\,$Mpc if $n_s \gtrsim 10^{-7}\,\rm Mpc^{-3}$. However, if we achieve $N_{\rm tot}^{\rm ev}\sim10^3$ with planned EeV neutrino detectors, it is also possible to find distant UHE neutrino emitters. If the sources are rare but powerful as expected in blazars, typical sources that dominate the neutrino sky may be distant rather than nearby. 
 
\section{Summary and Discussion}\label{sec:discussion}
We have investigated the requirements for a future EeV neutrino detector to identify a neutrino point source. We find that for non-evolving sources with $n_s\sim{10}^{-7}-{10}^{-5}~{\rm Mpc}^{-3}$, $\gtrsim100-1000$ events and sub-degree angular resolution are needed for a $\gtrsim5\sigma$ detection of UHE neutrino sources. This detection would also give relevant clues to the origins of UHECRs. The results are sensitive to the redshift evolution model, and the similar numbers are obtained for the SFR evolution with $n_s\sim{10}^{-9}-{10}^{-7}~{\rm Mpc}^{-3}$. By examining the typical distance to sources that can be identified by a UHE neutrino detector, we show that for source population with a number density above $\sim 10^{-6}\,\rm Mpc^{-3}$, a significant fraction of the brightest sources may be in the nearby Universe. Therefore if the sources are neither beamed nor transient, it would be possible to associate the detected sources with nearby objects observed using other messengers, including messengers with limited horizons. On the other hand, if sources are rare and powerful as predicted in blazar scenarios, they can be first found at distant locations. Note that UHECRs above the GZK energy should be suppressed for distant sources, but $\sim{10}^{19}$~eV cosmic rays may reach the Earth and their powerful sources may be relevant for anisotropy searches in the UHE range.  

So far we have only considered steady sources which therefore do not evolve over time. 
UHECRs from a transient event are expected to arrive at Earth with a spread of arrival times, which we designate by $\delta t$, because UHECR paths are deflected by magnetic fields between the source and the Earth. Therefore, for transient sources, the source rate per volume $\rho_{\rm s}$ can be converted to an apparent number density $n_{\rm s}$ via the UHECR arrival time spread: $\rho_{\rm s}\sim n_{\rm s}/\delta t$ \citep{Murase_Takami09,Takami:2011nn}.  
The time spread is $\delta t\sim 10^4\,{\rm yrs} \,(D/100 {\rm \,Mpc})\,(\delta\theta/1^\circ)^2$, where $\delta\theta$ is the angular deflection experienced by the particle during the propagation (e.g., \cite{KL08b}). The apparent number density may still govern the potential of pinpointing the sources of cosmogenic neutrinos, while the luminosity of EeV neutrinos from the sources can be much higher because the duration of UHE neutrino emission is short. 
Example transient sources of EeV neutrinos are GRB afterglows \cite{Waxman:1999ai,Dermer:2000yd,Murase:2007yt,Razzaque:2013dsa}, young magnetars and pulsars \cite{Murase09,FKMO14_letter}. Searches for transient UHE neutrino sources are relevant especially if the composition is dominated by protons and light nuclei. 

Unlike astrophysical UHE neutrinos, cosmogenic neutrinos are expected to suffer some, typically degree-level, angular displacement with respect to their source directions, due to the deflections suffered by the primary UHECRs that produce them. Therefore it may be difficult to pinpoint the actual source location even with a perfect detector.  On the other hand, our work suggests that once the detected number of cosmogenic neutrinos reaches a few hundred, crucial information can be gained regarding the characteristics of the sources of UHECRs. Importantly, any small scale auto-correlations in the arrival directions of cosmogenic neutrinos, or the absence thereof, would constrain the number density and source evolution of UHECR sources, in a measurement that is complementary to the equivalent study of UHECR arrival directions, and PeV neutrinos, as presented in e.g., Refs.~\cite{1997ApJ...483....1W, 2000PhRvL..85.1154D, AugerBound, 2016arXiv160701601M}.
 
In this work we only use the spatial information of neutrino events to look for sources. Therefore the significance of the searches does not depend on the energy spectrum of the events.  However, a different energy spectrum could impact the implications through two ways: 1) by changing  the expected total number of events (see equation~\ref{eq:Nev}), and 2) by changing the contribution of distant sources in a redshifted energy band. The influence of different spectral templates is demonstrated in Ref.~\cite{2016arXiv160701601M}.

The current work also applies to astrophysical neutrinos at lower energies including the energy range covered by IceCube.   Above $\gtrsim50-100$~TeV, coincidental pairs of muon neutrinos from the atmospheric background are negligible in most part of the sky \citep{2016arXiv160701601M}.  For lower-energy tracks or shower events with a poorer angular resolution,  a marginalization over energies or an energy-dependent likelihood~\citep{Fang:2016hyv} is necessary to avoid the confusion from the atmospheric events.  Our 90\% confidential level ($\sim1.6\sigma$ by the conventional conversion with a Gaussian) results shown in Figures~\ref{fig:sigma} and \ref{fig:sigma7} are consistent with the number density constraints presented by Ref.~\citep{2016arXiv160701601M} (note that the number of $\gtrsim60$~TeV neutrinos observed in the six-year observation is $\sim100$ in the half sky.). Our results in Figures~\ref{fig:sigma},\ref{fig:sigma7}, and \ref{fig:fracMultiplet} also predict the point-source search potential of future statistics from IceCube and ANTARES, as well as next-generation detectors like IceCube-Gen2 \citep{2014arXiv1412.5106I} and KM3NET \citep{2016JPhG...43h4001A}. 

\section*{Acknowledgements}
We thank Markus Ahlers and Olivier Martineau for helpful comments.  K. F. acknowledges the support of a Joint Space-Science Institute prize postdoctoral fellowship.  K. K. acknowledges financial support from the PER-SU fellowship at Sorbonne Universit\'es and from the Labex ILP (reference ANR-10-LABX-63, ANR-11-IDEX-0004-02).
The work of K. M. is supported by NSF Grant No. PHY-1620777.  We also acknowledge the University of Maryland supercomputing resources (http://www.it.umd.edu/hpcc) and  The Maryland Advanced Research Computing Center (https://www.marcc.jhu.edu) made available for conducting the research reported in this paper.   
\bibliography{EeVNeuDetection}

\end{document}